\documentstyle[12pt,epsfig]{article}
\begin{document}
\begin{flushright}
ADP-95-36/T190
\end{flushright}
\begin{center}
\begin{LARGE}
Effect of Nucleon Structure Variation \\
in Super-allowed Fermi Beta-decay
\end{LARGE}
\end{center}
\vspace{0.2cm}
\begin{center}
\begin{large}
K.~Saito~\footnote{ksaito@nucl.phys.tohoku.ac.jp} \\
Physics Division, Tohoku College of Pharmacy \\ Sendai 981, Japan \\
and \\
A.~W.~Thomas~\footnote{athomas@physics.adelaide.edu.au} \\
Department of Physics and Mathematical Physics \\
and \\
Institute for Theoretical Physics \\
University of Adelaide, 5005, Australia \\
\end{large}
\end{center}
%
\begin{abstract}
There is a well known anomaly between the value of the Fermi decay
constant extracted from super-allowed Fermi beta-decay of nuclear
isotriplets and that required by unitarity of the
Cabibbo-Kobayashi-Maskawa matrix. This discrepancy remains at the level
of a few tenths of a percent after the most rigorous investigation of
conventional nuclear and radiative corrections. Within the framework of
the quark-meson coupling model of nuclear matter, which has been previously
applied successfully to phenomena such as nuclear saturation and
nuclear charge symmetry violation, we show that it is possible to
understand a significant fraction of the observed anomaly.
\end{abstract}
\newpage

It is clearly very important to refine our understanding of the weak
coupling to quarks as much as possible. Testing the unitarity of the
Cabibbo-Kobayashi-Maskawa (CKM) matrix is one of the more challenging aspects
of this general problem. In particular, the most accurate
experimental measurement of
the vector coupling constant in nuclear beta-decay comes from
super-allowed $0^+$-$0^+$ transitions between nuclear isotriplet states.
However, in order to relate these precise measurements to the
quark-level vector coupling, $V_{ud}$, one needs to apply a number of
small nuclear structure corrections~\cite{rad} in addition to the relatively
standard radiative corrections~\cite{town}. Despite intensive study of
these nuclear ``mismatch'' corrections~\cite{mf1,mf2,hf,Rm} there
remains a systematic difference of a few tenths of a percent
between the value of $V_{ud}$ inferred
from the vector coupling measured in muon decay, $G_{\mu}$, and
unitarity of the CKM matrix and those determined from the nuclear
$ft$-values. For recent summaries we refer to the reviews of
Wilkinson~\cite{wilk} and Towner and Hardy~\cite{towner}, and also
to the recent report by Savard et al.~\cite{savard} of accurate
data on $^{10}$C.

Until now the nuclear corrections have been explored within the
framework of conventional nuclear theory with point-like nucleons.
Of course, for the nucleon itself there has been considerable
investigation of the effect on the vector form-factor of the breaking of
CVC caused by the small $u$-$d$ mass difference in QCD. The Ademollo-Gatto
theorem~\cite{ad,bs} tells us that any corrections must be of second
order in $(m_d - m_u)$ -- a result that has survived~\cite{dw,pr}
suggestions that it might fail because of $\rho$-$\omega$ mixing~\cite{cp}.
While this is necessarily very small, the measurements of $V_{ud}$ and
$G_{\mu}$ are also extremely precise. Thus we have been led to ask
whether this small nuclear discrepancy might be associated with a change
in the degree of non-conservation of the vector current
caused by nuclear binding.

In order to investigate whether nuclear binding might influence the Fermi
decay constant of the nucleon itself one needs a model of nuclear
structure involving explicit quark degrees of freedom which nevertheless
provides an acceptable description of nuclear binding and saturation.
The quark-meson coupling (QMC) model of Guichon~\cite{guichon} seems
ideally suited to the problem. In this
model, nuclear matter consists of non-overlapping nucleon
bags bound by the self-consistent exchange of $\sigma$ and $\omega$
mesons in the mean-field approximation.
It has been extended by Yazaki et al.~\cite{yazaki} to include a centre
of mass correction and by the present authors to include the $\rho$ and
an isovector-scalar meson (the
$\delta$)~\cite{sath,ons}. As well as providing an excellent description
of the properties of nuclear matter,
it has been applied successfully to the
calculation of nuclear structure functions~\cite{saito} and (allowing
for a difference in the quark masses, $m_u \neq m_d$) to the
problem of the Okamoto-Nolen-Schiffer anomaly~\cite{okns} in mirror
nuclei~\cite{ons}.
Furthermore, the relationship between
the QMC model and Quantum Hadrodynamics~\cite{serot} has been
investigated.  The fascinating result is that for infinite nuclear matter
the two approaches can be written in an identical form, except for
the appearance of the quark-scalar density in the self-consistency
condition for the scalar field~\cite{sath,comp}. The simplicity of
this finding suggests that it may be rather more general than the
specific model within which it was derived.

We shall briefly review the main features of the model with unequal
quark masses before turning to the specific problem of the Fermi
form-factor. Suppose that
the mean-field values for the $\sigma$, $\omega$ (the time
component), $\rho$ (the time component in the third direction of
isospin) and $\delta$ (in the third direction of isospin) fields,
in uniformly distributed nuclear matter with $N \ne Z$, are
$\bar{\sigma}$, $\bar{\omega}$, $\bar{b}$ and $\bar{\delta}$,
respectively.  The
nucleon is described by the static spherical MIT bag in
which quarks interact (self-consistently)
with those mean fields.  Then the Dirac equation for
a quark field, $\psi$, in a bag is given by
\begin{equation}
[i\gamma\cdot\partial - (m_i - V_{\sigma} - \frac{1}{2}
\tau_zV_{\delta}) - \gamma^0(V_{\omega} + \frac{1}{2}\tau_zV_{\rho})]
\psi_{i/j} = 0, \label{eq:dirac}
\end{equation}
where $V_{\sigma}=g_{\sigma}^{q}\bar{\sigma}$, $V_{\omega}=
g_{\omega}^{q}\bar{\omega}$, $V_{\rho}=g_{\rho}^q\bar{b}$ and
$V_{\delta}=g_{\delta}^q\bar{\delta}$ with the quark-meson coupling
constants, $g_{\sigma}^q$, $g_{\omega}^q$, $g_{\rho}^q$ and
$g_{\delta}^q$.  The subscripts, {\it i\/} and {\it j\/}, denote
the {\it i}-th quark ({\it i\/}={\it u\/} or {\it d\/})
in the proton or neutron ({\it j\/}={\it p\/}
or {\it n\/}).

The normalized, ground state wave function
for a quark in the nucleon is then given by
\begin{equation}
\psi_{i/j}(\vec{r},t) = {\cal N}_{i/j} \exp[-i\epsilon_{i/j} t/R_j]
{j_{0}(x_ir/R_j) \choose
i\beta_{i/j} {\vec{\sigma}}\cdot\hat{r}j_{1}(x_ir/R_j)}
{\frac{\chi_i}{\sqrt{4\pi}}}, \label{eq:psiq}
\end{equation}
where
\begin{equation}
\epsilon_{i/j} = \Omega_{i/j} + R_j(V_{\omega} \pm \frac{1}{2}
V_{\rho}),
\mbox{ for a } {u \choose d} \mbox{ quark} \label{eq:epq}
\end{equation}
\begin{equation}
{\cal N}_{i/j}^{-2} = 2R_j^3j^2_0(x_i)[\Omega_{i/j}(\Omega_{i/j} - 1)
+
R_jm_i^{\star}/2]/x_i^2, \label{eq:norm}
\end{equation}
\begin{equation}
\beta_{i/j} = \sqrt{(\Omega_{i/j} - R_jm_i^{\star})/(\Omega_{i/j} +
R_jm_i^{\star})}, \label{eq:betq}
\end{equation}
with $\Omega_{i/j} = \sqrt{x_i^2 + (R_jm_i^{\star})^2}$ and $\chi_i$
the
quark spinor.  The effective quark mass, $m_i^{\star}$, is defined
by
\begin{equation}
m_i^{\star} = m_i - (V_{\sigma} \pm \frac{1}{2}V_{\delta}),
\mbox{ for a } {u \choose d} \mbox{quark}. \label{eq:qem}
\end{equation}
The linear boundary condition at the bag
surface determines the eigenvalue $x_i$.

Taking the spin-flavor wave function for the nucleon to have the
usual SU(6) form,
the nucleon energy is given by $E_{bag}^j + 3V_{\omega} \pm \frac{1}
{2}V_{\rho}$ for ${p \choose n }$, where the bag energy is
\begin{equation}
E_{bag}^j = {\frac{\sum_i n_{i/j} \Omega_{i/j} - z_0}{R_j}} +
{\frac{4}{3}}\pi
BR_j^3,
\label{eq:bageb}
\end{equation}
$B$ is the bag constant and $z_0$ is a phenomenological parameter
initially introduced to account for zero-point motion.
Here we use it to correct for spurious c.m. motion as well -- rather
than following the more elaborate procedure of Ref.~\cite{yazaki}. The
reason is that, as illustrated by the Ademollo-Gatto theorem, the
deviation of the vector form-factor from unity is a very subtle effect.
This method is the only one in which we have been able to
guarantee that in the free nucleon case the
deviation is proportional to $(m_d - m_u)^2$.
The effective nucleon mass, $M_j^{\star}$, in nuclear matter is given
by minimizing eq.(\ref{eq:bageb}) with respect to $R_j$.

To see the sensitivity of our results to the bag radius of the free
nucleon, we choose $m_u=5$ MeV and vary the parameters, $B$, $z_0$ and
$m_d$, to fit the physical proton and neutron masses for several
values of the average, free bag radius $R_0$ (= 0.6, 0.8, 1.0 fm).
Since the
electromagnetic (EM) self-energies for {\it p\/} and {\it n\/}
contribute to the masses we adjust the parameters to fit the bare
proton mass, $M_p = 938.272-0.63$ MeV, and the bare neutron mass,
$M_n=939.566+0.13$ MeV, where $+0.63$ MeV and $-0.13$ MeV are the
EM self-energies for {\it p\/} and {\it n\/},
respectively~\cite{elms}.
We then find that $B^{1/4}$ = (210.7,
169.8, 143.6) MeV, $z_0$ = (4.011, 3.305, 2.600), and $m_d$ = (9.252,
9.242, 9.233) MeV for $R_0$ = (0.6, 0.8, 1.0) fm, respectively.
It is crucial for the present application that the
bag radius of the proton is a little smaller than that of the
neutron in matter and that this difference increases slightly with density.

At the present time the extension of the QMC model to finite nuclei is
not complete~\cite{prep}.
In particular, the variation of the scalar and vector
fields across the finite size of the nucleon presents quite a challenge.
Thus, for the present we are constrained to discuss the problem within
the framework of infinite nuclear matter. In this case
we take the Fermi momenta for protons and
neutrons to be $k_{F_p}$ and $k_{F_n}$, respectively. These are
defined by $\rho_p = k_{F_p}^3 / (3\pi^2)$ and $\rho_n = k_{F_n}^3 /
(3\pi^2)$, where $\rho_p$ and $\rho_n$ are the densities of
{\it p\/} and {\it n}, respectively, and the total baryon density,
$\rho_B$, is given by $\rho_p + \rho_n$.  The $\omega$ field is
determined by baryon number conservation,
and the $\rho$ mean-field by the difference in proton and neutron
densities ($\rho_3$ below). On the other hand, the scalar mean-fields,
${\bar \sigma}$ and ${\bar \delta}$, are given by self-consistency
conditions (SCCs)~\cite{sath,ons}.  Since the $\rho$ field value is
given by $\bar{b} = g_{\rho} \rho_3 / (2m_{\rho}^2)$, where
$g_{\rho}=g_{\rho}^q$ and $\rho_3 = \rho_p - \rho_n$, the total
energy per nucleon, $E_{tot}$, can be written
\begin{equation}
E_{tot} = \frac{2}{\rho_B (2\pi)^3}\sum_{j=p,n}\int^{k_{F_j}}
d\vec{k} \sqrt{M_j^{\star 2} + \vec{k}^2} + \frac{m_{\sigma}^2}
{2\rho_B}{\bar{\sigma}}^2 + \frac{m_{\delta}^2}{2\rho_B}
{\bar{\delta}}^2 + \frac{g_{\omega}^2}
{2m_{\omega}^2}\rho_B + \frac{g_{\rho}^2}{8m_{\rho}^2\rho_B}
 \rho_3^2, \label{eq:toteb}
\end{equation}
where $g_{\omega} = 3g_{\omega}^q$.

We determine the coupling constants, $g_{\sigma}^2/4 \pi$ =
(22.62, 20.78, 19.47), where $g_{\sigma} = 3 g_{\sigma}^q$, and
$g_{\omega}^2/4\pi$ = (5.547, 4.637, 3.979) for $R_0$ = (0.6, 0.8, 1.0) fm,
respectively, so as to fit the binding
energy ($-16$ MeV) and
the saturation density ($\rho_0$ = 0.17 fm$^{-3}$) for equilibrium
nuclear matter.
It is important to stress that the $\delta$ plays an insignificant role in
our present calculation, while the $\rho$ has no influence in symmetric
matter. Nevertheless, for the record we note that
$g_{\delta}^2/4\pi$ is chosen to be
2.82~\cite{bonn} (where $g_{\delta} = g_{\delta}^q$) and the $\rho$
meson coupling constant,
$g_{\rho}^2/4\pi$, is chosen to be (5.223, 5.340, 5.423)
for $R_0$ = (0.6, 0.8, 1.0) fm, respectively,
so as
to reproduce the bulk symmetry energy of nuclear matter, $33.2$ MeV.
One of the important consequences of including the internal structure of
the nucleon is that the model gives a very respectable
value for the nuclear incompressibility, around 200 $\sim$ 300 MeV.

Armed with an acceptable, quark-based model of nuclear matter we can now
investigate the variation with density of the quark vector current
matrix element. For Fermi beta-decay we need the matrix elements
\begin{equation}
I_{ii'}(\rho_B) = \int_{Bag} dV \psi_{i/p}^{\dagger} \psi_{i'/n},
\label{Iji}
\end{equation}
with $i'=d$ and $i=u$ for the $d \rightarrow u$ conversion and $i=i'=u$ or
$d$ for the two spectator quarks. As the radius of the proton and
neutron are different we integrate over the common volume.
\begin{figure}[tb]
\begin{center}
\epsfig{file=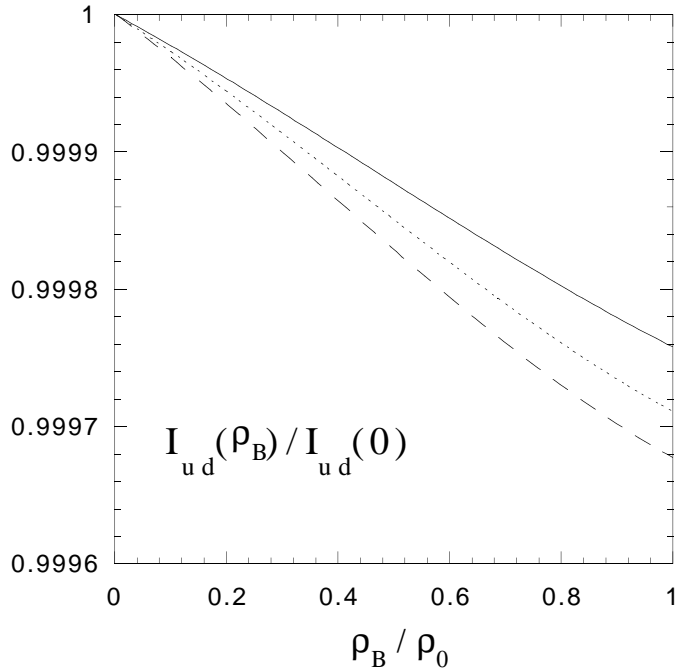,height=9cm}
\caption{$I_{ud}(\rho_B)/I_{ud}(0)$ as a function
of the nuclear density.
The solid, dotted and dashed
curves are for $R_0$ = 0.6, 0.8 and 1.0 fm,
respectively. }
\label{fig:density}
\end{center}
\end{figure}

As remarked earlier, the first test of the consistency of the
calculation is that at zero density the deviation of the vector
form-factor from unity is quadratic in $m_d - m_u$.
By a numerical check we can see that this is indeed the case .

In fig.\ref{fig:density} we show the ratio
$I_{ud}(\rho_B)/I_{ud}(0)$, for our
three choices of the free (average) bag radius (in symmetric
nuclear matter), as a function of
$\rho_B$.  We also give full details of those values at several nuclear
densities in table~\ref{tbl}.
The decrease in $I_{ii'}$ as
the density increases is a direct consequence of the increasing
difference between the proton and neutron radii -- that is the smaller
volume of overlap.
\begin{table}[hbtp]
\begin{center}
\caption{Variations with density of $I_{ii'}$ for three bag radii.}
\label{tbl}
\begin{tabular}[t]{ccccc}
\hline
$R_0$(fm) & $\rho_B/\rho_0$ & $I_{ud}$ & $I_{uu}$ & $I_{dd}$ \\
\hline
 &     0  & 0.999993 & 0.999995 & 0.999995 \\
0.6 & 0.5 & 0.999869 & 0.999870 & 0.999871 \\
 &    1.0 & 0.999751 & 0.999751 & 0.999752 \\
\hline
 &     0  & 0.999991 & 0.999994 & 0.999994 \\
0.8 & 0.5 & 0.999840 & 0.999842 & 0.999843 \\
 &    1.0 & 0.999702 & 0.999703 & 0.999705 \\
\hline
 &     0  & 0.999987 & 0.999992 & 0.999992 \\
1.0 & 0.5 & 0.999814 & 0.999817 & 0.999819 \\
 &    1.0 & 0.999664 & 0.999666 & 0.999669 \\
\hline
\end{tabular}
\end{center}
\end{table}
{}From the figure and the table, as the density
increases from zero, we see that the deviation of $I_{ii'}(\rho_B)/
I_{ii'}(0)$ from unity
is roughly linear in
$\rho_B$. (The deviation is also roughly linear in $m_d - m_u$.)
We can then summarise the results as:
\begin{equation}
\frac{I_{ii'}(\rho_B)}{I_{ii'}(0)} \simeq 1 - a_{ii'} \times
\left( \frac{\rho_B}{\rho_0}
\right), \label{Iii}
\end{equation}
with $a_{ii'} \simeq (2.4, 2.9, 3.3) \times 10^{-4}$ for $R_0$ =
(0.6, 0.8, 1.0) fm, respectively (for any combination of $ii'$).

The evaluation of $ft$-values involves the inverse of the product of
$I_{ud}, I_{uu}$ and $I_{dd}$ squared.  Since for a given,
free (average) radius
of the bag each of these matrix elements
decreases by roughly the same amount, the fractional
increase in the $ft$-value with
density is therefore
\begin{equation}
\frac{ft(\rho_B)}{ft(0)} \simeq 1 + b \times
\left( \frac{\rho_B}{\rho_0}
\right), \label{ft}
\end{equation}
with $b$ approximately six times the decrease in each
integral -- i.e. $b \simeq (1.5, 1.8, 2.0) \times
10^{-3}$ for $R_0$ = (0.6, 0.8, 1.0) fm.
Thus the increase in the $ft$-value at $\rho_0/2$ ranges from $0.075\%$
to $0.10\%$, while at $\rho_0$ it lies between $0.15\%$ and $0.20\%$.
This is to be compared with a violation of unitarity of the CKM matrix
of $0.35 \pm 0.15\%$ in the most recent
analysis of Towner and Hardy~\cite{towner}.

It is not possible to
draw unambiguous conclusions from a comparison of
theoretical results in infinite nuclear matter with data from finite
nuclei. At $\rho_0/2$ the calculation suggests a reduction in the
violation of unitarity by about $1/3$, while at $\rho_0$ a correction as
big as $0.2\%$ brings the discrepancy back to only one standard
deviation.  We also note that this result is rather sensitive to
the proton fraction, $f_p = \rho_p/\rho_B$.  The overlap
integrals with $f_p$ less (greater) than 0.5
decrease more slowly (rapidly) than in the symmetric case.

Before concluding our discussion it might be helpful to make a
qualitative observation concerning the validity of our results.
The essential physics involved in our calculation is charge symmetry
violation, in particular, the fact that {\it in nuclear matter}
the confining potential felt by a
quark in a proton is not the same as that felt by a quark
in a neutron.
We have already  explained that a relativistic field theory only yields the
right order of magnitude for nuclear charge symmetry breaking if the
relevant mass scale involves quarks rather than nucleons~\cite{julich}.
In this sense the ONS anomaly may prove to
be something of a ``smoking gun'' for
quark degrees of freedom in nuclei. This is even more obvious here; it
is only because the nuclear charge symmetry violation occurs at the
quark level that it can produce a deviation of the vector form factor of
the bound nucleon from its free value.

In summary, within a specific, relativistic quark-meson coupling model
of nuclear
matter we have found that the change of the internal structure of the
nucleon itself produces a correction to the Fermi beta-decay form-factor
of the sign and order of
magnitude required for consistency with the unitarity of
the Cabibbo-Kobayashi-Maskawa matrix.
It is important that the model
dependence of this result be further examined as soon as possible
-- although we
have given arguments why the qualitative result may be quite general.
Finally we note that the present model deals only with nuclear matter
and it is vital that the formalism be extended so that we can apply the
model quantitatively in finite nuclei.

We would like to thank Peter Jackson for raising this issue with one of
us during the celebration of Erich Vogt's 65th birthday
at TRIUMF last December. We would also like to acknowledge helpful
discussions with T. Hatsuda, I. Towner and D. Wilkinson during WEIN'95.
This work was supported by the Australian Research Council.

%
%

%
\end{document}